\documentclass[12pt]{article}
\usepackage{psfig}
\usepackage{graphicx}

\begin{document}

{Astronomy Letters, Vol. 32, No. 5, pp. 302-307, 2006} \\ \\ \\ \\

\begin{center}

\bigskip

{\LARGE \bf Interacting Galaxies in Sparsely Populated Groups} \\

{\small O. V. Melnyk \footnote {O. V. Mel'nik}}
 \\

{\em Astronomical Observatory, Kiev National University, 3 Observatorna str.,
Kiev, 04053 Ukraine \\
melnykov@observ.univ.kiev.ua} \\

\end{center}

{\large \bf Abstract}\\
We calculated the median values of the following parameters for 87 groups of galaxies with three
to eight components: the mean rms velocity of the galaxies in the group, $S_{v}$ = 166 km/s; the harmonic
mean radius, $R_{h}$ = 29 kpc; and the mass-to-light ratio, $M_{vir}/L = 33 M_{\odot}/L_{\odot}$. The $M_{vir}/L$ ratio depends
on the population of the system, while $S_{v}$ does not depend on $R_{h}$. To ascertain the relationship between
the activity of galaxies in groups and their morphological composition and the effect of the environment
on the evolutionary processes in groups, we consider the fraction of galaxies with UV excess in the
sample of interacting galaxies in groups (6\%), single peculiar galaxies (8\%), and isolated galaxies (4\%)
and their morphological composition. We also show that the number of active galaxies decreases with
increasing population of the group of galaxies, while the frequency of occurrence of early-type ($E/S0$)
galaxies increases.

\bigskip

\section{Introduction}
Vorontsov-Velyaminov, Arp, and other authors began an active discussion of the nature of interacting galaxies in the 1950s. The definition of interacting galaxies as objects embedded in a common luminous shell, interpenetrating, and with distortions of their shapes (asymmetry, tails, and bridges) belongs to Vorontsov-Velyaminov (1958). Photographic, spectroscopic, and photometric observations of interacting galaxies were carried out in the 1970s at the Special Astrophysical Observatory of the Russian Academy of Sciences and the Crimean and Byurakan Astrophysical Observatories. Interacting galaxies that form groups or are components of compact groups of galaxies (see, e.g., Karachentseva and Karachentsev 2000, Karachentseva et al. 1987, Hickson et al. 1989) clusters of galaxies, are encountered; there are also isolated peculiar objects among interacting galaxies (Karachentseva 1973). Compact Hickson groups (Hickson et al. 1989) that contain a significant fraction of interacting galaxies enjoy the greatest popularity from both observational
and theoretical points of view. 

Studying the relationship between the activity of galaxies in groups and their morphological composition is important for understanding the evolutionary processes in close groups. Comparison of the kinematic parameters and mass-to-light ratios for groups containing different numbers of galaxies with those for isolated galaxies provides no less important information. To this end, we compare the occurrence of active galaxies with UV excess in groups with different numbers of members containing interacting galaxies from the catalog by Vorontsov-Velyaminov et al. (2001) and in the sample of isolated galaxies by Karachentseva (1973). In this paper, we also consider observational, kinematic, and virial properties of these systems.

The catalog of interacting galaxies (Vorontsov-Velyaminov 2001) is a combination and continuation of the two previous parts of the catalog published by Vorontsov-Velyaminov in 1959 and 1977. The catalog contains 2014 interacting galaxies and is homogeneous to $15^{m}$ and $\delta > -45^{\circ}$. However, it also includes faint galaxies (to $20^{m}$); the radial velocities of the objects reach 27 000 km/s. Among the 2014 interacting galaxies of the catalog, 446 (22\%) and 73 (3\%) were classified as nests N and chains Ch, respectively; pairs P are most numerous, 1230 (61\%);
and the remaining 14\% are peculiar galaxies or M51-type interacting galaxies with close companions. \\ \\

\section{The Sample}

Systems classified by the type of interaction as nests N or chains Ch with three or more members (2/3 and 1/3 of the systems in the sample contain nests and chains of galaxies, respectively) were selected from the catalog of Vorontsov-Velyaminov et al. (2001). Close galaxies, but without any traces of distortions of their shapes were not considered to be interacting galaxies according to the criterion of Vorontsov-Velyaminov (1958) and, accordingly, were not included in the catalog. Therefore, to identify a "bound" group of galaxies, we corrected the number of components in the system by visually checking each system on DSS images and galaxy parameters in the LEDA and NED databases. This selection was necessary, since, according to Zasov and Arkhipova (2000), more than a third of the nests of galaxies identified by Vorontsov-Velyaminov are actually single galaxies with several sites of star formation or pairs. The triplet VV143 (HCG18), which is actually a giant irregular galaxy (Plana et al. 2000), is a dramatic example. Thus, if an interacting system belonged to a more densely populated Hickson group, then the components were included in the group; conversely, if a galaxy with a radial velocity differing significantly from the radial velocities of other components was detected in the system, then it was excluded from the group (as, e.g., in the Stephan Quintet VV288). The sample contains a total of 30 groups identified with compact Hickson groups. The final sample contains 335 galaxies in 87 groups consisting of three to eight components with known observational parameters (radial velocities, sizes, apparent magnitudes, and morphological types) for all components. \\ \\

\section{The morphological composition and activity of interacting galaxies}

We compared the parameters of interacting galaxies in groups with two samples of galaxies. One of these samples consists of single peculiar galaxies that were classified in the catalog by Vorontsov-Velyaminov (2001) as En (enigmatic, peculiar), K (comet-like), H (with vast HII regions), and R (with rings); nests N or chains Ch, which may be barred spirals or irregular galaxies with large bright knots (Zasov and Arkhipova 2000); and M-galaxies with a satellite in the spiral arms. There are a total of 533 such galaxies in this sample; the radial velocities and apparent magnitudes are known for 471 of them. For comparison, we also selected 1022 from the 1051 galaxies of the catalog of isolated galaxies by Karachentseva (1973) (CIG), which have the status
of isolated galaxies in the NED database; the radial velocities, apparent magnitudes, and morphological types are known for 878 of them.

Figures 1 and 2 show the distribution of galaxies in morphological type for the sample of interacting galaxies in the selected groups and the comparison samples and their differential luminosity functions.

\begin{figure}[t]
\begin{tabular}{ll}
\includegraphics[angle=0, width=0.48\textwidth]{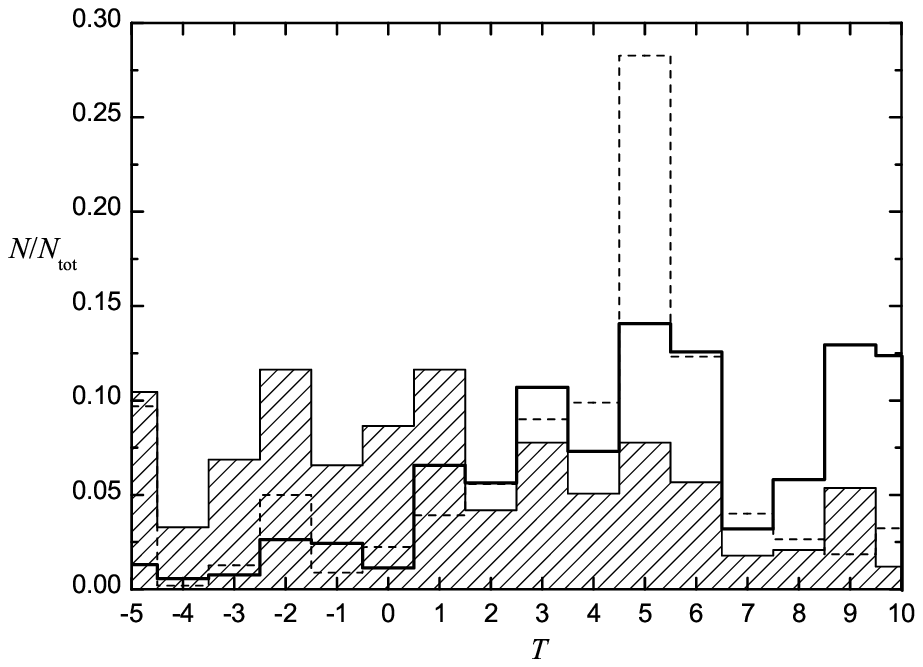} &
\includegraphics[angle=0, width=0.45\textwidth]{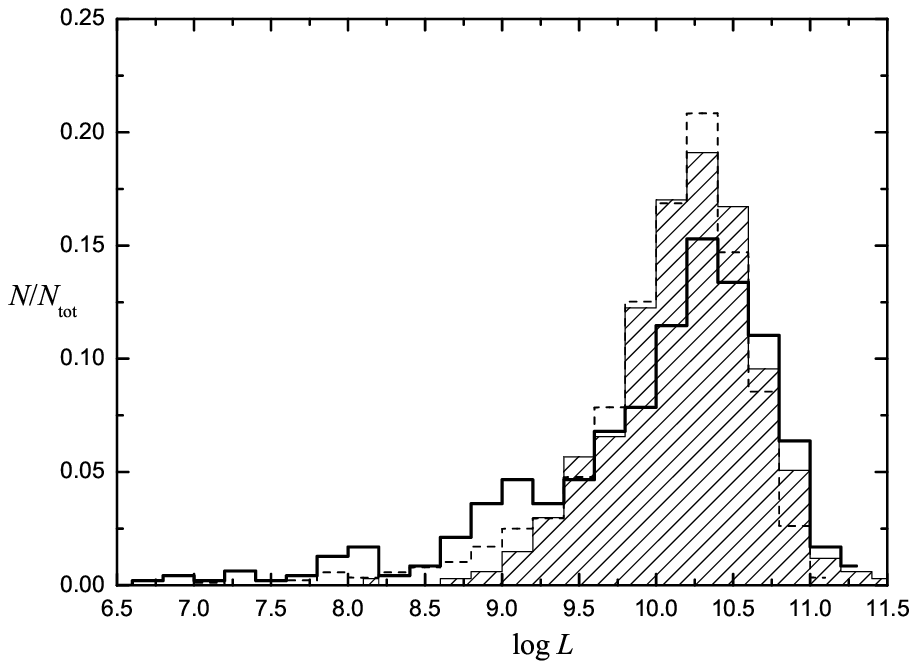} 
\end{tabular}
\caption{ (left) Distribution of galaxies in morphological type: the shaded diagram shows interacting galaxies in groups, the diagram marked by the solid line shows peculiar galaxies, and the diagram marked by the dashed line shows isolated CIG galaxies.}
\caption{ (right) Differential luminosity functions. The notation is the same as that in Fig. 1.}
\end{figure}

The fraction of early-type $E/S0$ (the LEDA classification -5..0) galaxies among the interacting galaxies in groups is 47\%. The fractions of early-type galaxies among the single peculiar and isolated CIG galaxies are 7\% and 19\%, respectively. The sample of single peculiar galaxies has more late-type and irregular galaxies and more low-luminosity galaxies.

The morphological type is luminosity-independent in all three samples, i.e., galaxies of various morphological types are represented in the entire range of luminosities.

Figure 3 shows the distribution of galaxies in ($U - B$) and ($B - V$) color indices (according to LEDA). The number of galaxies with known color indices in each sample is $N_{(U - B)}$ = 30 and $N_{(B - V)}$ = 41 for interacting galaxies in groups, $N_{(U - B)}$ = 77 and $N_{(B - V)}$ = 109 for single peculiar galaxies, and $N_{(U - B)}$ = 66 and $N_{(B - V)}$ = 92 for isolated CIG galaxies. The availability of a color index for a galaxy does not depend on its luminosity and morphological type.

\begin{figure}[t]
\begin{tabular}{ll}
\includegraphics[angle=0, width=0.5\textwidth]{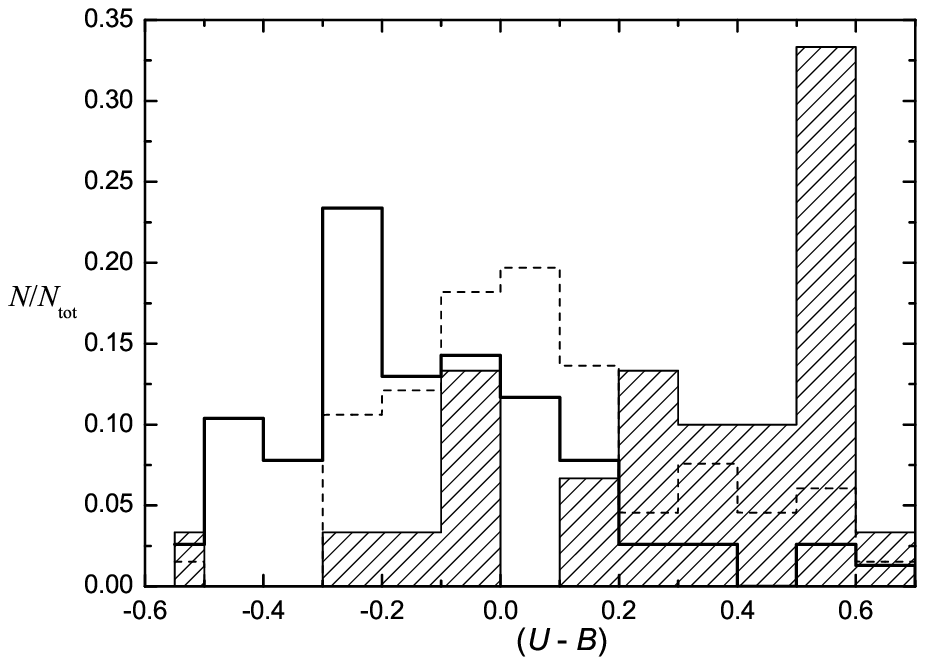} &
\includegraphics[angle=0, width=0.5\textwidth]{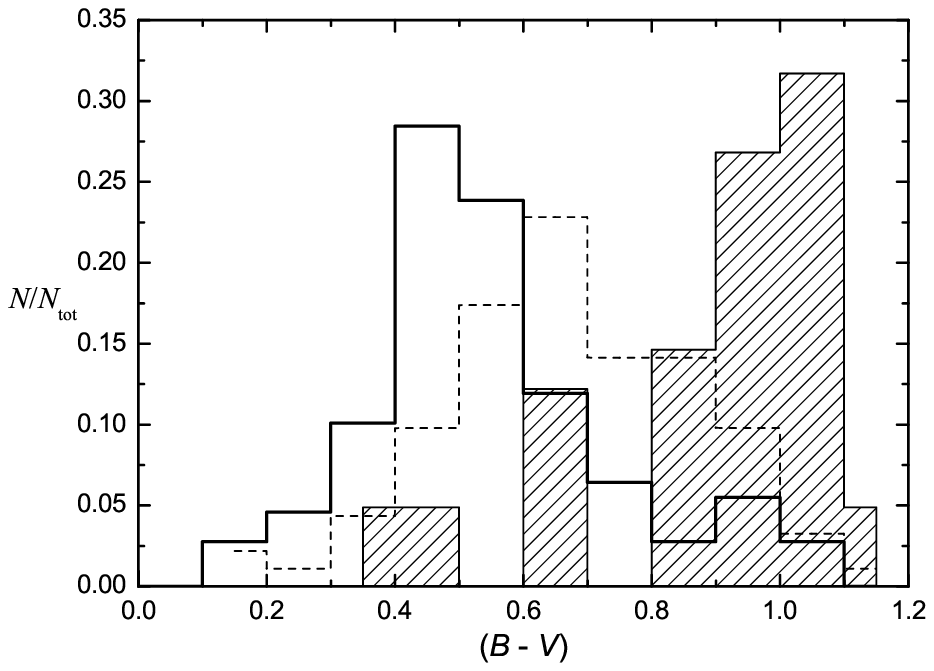} \\
\end{tabular}
\caption{Distribution of galaxies in $(U - B)$ (left) and $(B - V )$ (right) color indices. The notation is the same as that in Fig. 1.}
\end{figure}

Because of the poor statistics, we cannot analyze in detail the differences in color index, for example, for the same morphological type depending on the galaxy membership in a certain sample. However, it follows from Fig. 3 that the interacting galaxies in groups are redder than the isolated galaxies and considerably redder than the peculiar galaxies. Thus, the $(U - B)$ and $(B - V)$ color distributions for the two samples from the catalog by Vorontsov-Velyaminov (2001) confirm the performed morphological classification: an excess of early-type objects, i.e., galaxies of an older population among the interacting galaxies in groups, and an excess of irregular, blue objects among the single peculiar galaxies. 

We then compared the fractions of Markarian and Kazarian galaxies with UV excess (according to LEDA and NED) in the samples under consideration.We found that active galaxies account for 6\%, 8\%, and 4\% of the interacting galaxies in groups, single peculiar galaxies, and the CIG sample, respectively. The fractions of active galaxies among groups with different numbers of members are 9\%, 7\%, and 3\% for triplets, quartets, and groups with five to eight galaxies, respectively. Active galaxies with UV excess are encountered rarely among the early-type objects in all samples, 2\% for interacting galaxies in groups and $\sim$  1\% among single-galaxy samples. Therefore, most of the active objects in all samples are Sa..Irr galaxies. Galaxies with emission-line spectra, Seyferts, and LINERs are the dominant types of activity in all samples; galaxies with vast HII regions, active star formation, and absorption-line spectra are also encountered. We found no dominance of these activity types when comparing the isolated and interacting galaxies. The activity of a galaxy can be produced by both internal and external processes: the fall of a companion to the galaxy and tidal phenomena (see Byrd and Valtonen (2001) and references therein). The high percentage of active galaxies among the single peculiar galaxies and their occurrence among the isolated CIG galaxies could be the result of both factors.

According to Karachentsev and Karachentseva (1974) and Byrd and Valtonen (2001), there is an excess of galaxy pairs in which both components
are Markarian galaxies. We considered the fractions of active Markarian and Kazarian galaxies in the galaxy triplets of two samples: interacting triplets
(ITs) and the combined sample of Northern and Southern triplets (NSTs) (Karachentseva et al. 1987; Karachentseva and Karachetnsev 2000); the rms
velocity of the member galaxies is $S_{v} < 300$ km/s for all triplets. Note that the IT groups with $S_{v} > 300$ km/s contain no active galaxies with UV
excess.

The columns in Table 1 list the expected (E) (for a random distribution) and observed (O) numbers of triplets, $N_{tr}$, without galaxies with UV excess, with one UV galaxy, with two UV galaxies, and with three UV galaxies for the IT and NST samples. \\

Table 1. Number of triplets with UV galaxies.

\begin{center}
\begin{tabular}{|c|c|c|c|c|c|c|c|c|c|}
\hline
Sample & $N_{tr}$  & \multicolumn{2}{|c|}{$N_{tr}$ without UV} & \multicolumn{2}{|c|}{$N_{tr}$ with 1 UV} & \multicolumn{2}{|c|}{$N_{tr}$ with 2 UV} & \multicolumn{2}{|c|}{$N_{tr}$ with 3 UV} \\
\cline{2-10}
 & total & E & O & E & O & E & O & E & O \\
\hline
IT & 42 & 30 $\pm$ 5  & 32 & 11 $\pm$  3 & 7 & 1 $\pm$ 1 & 2 & 0 & 1\\
\hline
NST & 86 & 79 $\pm$ 9 & 78 & 10 $\pm$ 3 & 6 & 0 & 1 & 0 & 1\\
\hline
\end{tabular}
\end{center}

It can be seen from Table 1 that the expected number of triplets for a random distribution without active galaxies and with two UV galaxies is smaller
than their observed number in both samples, although the difference is within the error limits. The expected number of triplets with one UV galaxy is greater than their observed number, while the expected number of triplets with three UV galaxies is smaller than their observed number. Three galaxies with UV excess have two triplets: VV672 and KTG82 (the VV2002 pair). Thus, among the triplets, as in double galaxies (Karachetnsev and Karachentseva 1974; Byrd and Valtonent 2001), there is an excess of groups where all components are active galaxies with UV excess. 

The occurrence of active galaxies in interacting groups is also related to their population: the more components in the group, the fewer galaxies with UV excess it contains (the fraction of active galaxies is only 3\% in groups of five to eight members). These data may be indicative of different evolutionary stages of these systems. Coziol et al. (2004) suggested an evolutionary theory of compact groups based on the morphological type and spectral indices of galaxy activity determined for 27 compact Hickson groups of galaxies. The evolved groups, i.e., those in which a merger has already occurred, contain elliptical galaxies that exhibit no activity, while at the first merger stage, there are late-type galaxies with active star formation in the groups.

\section{Kinematic and virial parameters}

We calculated the kinematic and virial parameters for 87 groups in the same way as was done by Karachentseva and Karachentsev (2000) and Vavilova
et al. (2005) for groups with different numbers of galaxies. Table 2 lists the median parameters: $n_{m}$ is the number of galaxies in the group; $N_{gr}$ is the number of groups; $N(E/S0)$ is the ratio of the number of early-type galaxies to the total number of galaxies in the group: $N(E/S0)$ = $n(E/S0)/n_{m}$; $\langle V_{LG} \rangle $ are the mean radial velocities of the groups corrected for the solar motion as prescribed by Karachentsev and Makarov (1996), in km/s; $S_{v}$ are the rms velocities of the galaxies, in km/s; $R_{h}$ is the harmonic mean radius of the group, in kpc; $\tau$ is the dimensionless crossing time of the system, in units of the Hubble time 1/$H_{0}$; $M_{vir}$ is the virial mass of the group, in $10^{12}$ $ÌM_{\odot}$; $L$ is the luminosity, in $10^{10}$ $L_{\odot}$; and $M_{vir}/L$ is the virial mass-to-light ratio, in $M_{\odot}/L_{\odot}$. 

Table 2. Median kinematic and virial parameters of groups.

\begin{tabular}{|c|c|c|c|c|c|c|c|c|c|}
\hline
$n_{m}$ & $N_{gr}$ & $N(E/S0)$ & $\langle V_{LG} \rangle$,  & $S_{v}$, & $R_{h}$,  & $\tau$,  & $M_{vir}$,  & $L$,  & $M_{vir}/L$,  \\

& & & km/s &  km/s & kpc & 1/$H_{0}$ & $10^{12}$ $M_{\odot}$ & $10^{10}$ $L_{\odot}$ & $M_{\odot}/L_{\odot}$ \\
\hline
3 & 51 & 0.33 & 7291 & 104 & 26 & 0.04 & 0.99 & 6.16 & 19\\
\hline
4 & 19 & 0.50 & 7265 & 237 & 38 & 0.02 & 5.33 & 8.52 & 53\\
\hline
5-6 & 10 & 0.78 & 8628 & 385 & 33 & 0.02 & 11.8 & 9.21 & 99\\
\hline
7-8 & 7 & 0.55 & 9004 & 384 & 60 & 0.02 & 24.6 & 13.3 & 151\\
\hline
3-8 & 87 & 0.43 & 7673 & 166 & 29 & 0.03 & 2.29 & 7.53 & 33\\
\hline
\end{tabular}\\ \\

\begin{figure}[ht]
\centerline{\includegraphics[angle=0, width=10cm]{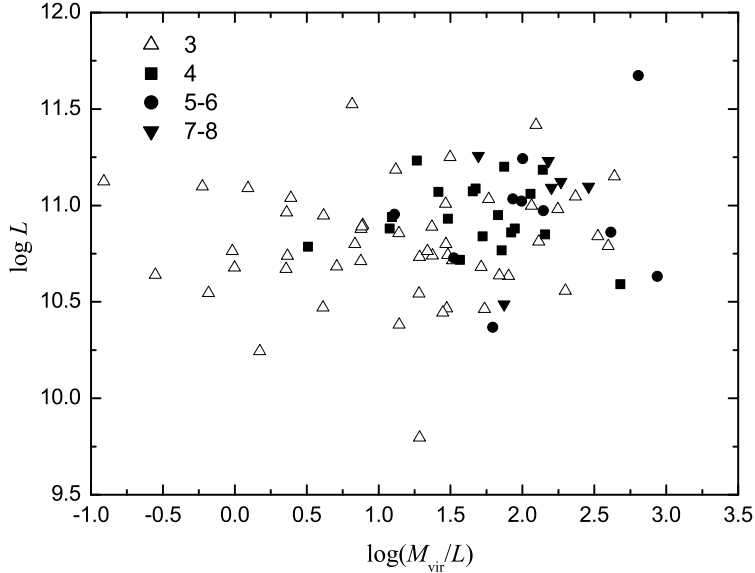}}
\caption{Luminosity vs. virial mass-to-light ratio for variously populated groups of galaxies.}
\end{figure}

It follows from Table 2 that the velocity dispersion of the group and its mass, luminosity, and virial mass-to-light ratio increase with increasing group population. The mass-to-light ratio $M_{vir}/L$ depends strongly on the rms velocity and does not depend on the harmonic mean radius. Figure 4 shows that the luminosity of a group increases only slightly with $M_{vir}/L$. 

Triplets are found in the entire $M_{vir}/L$ range; the spread in $M_{vir}/L$ is greatest for them. Almost all of the triplets located in Fig. 4 to the right in the region of groups consisting of four to eight components have rms velocities higher than 300 km/s. In addition, such high $S_{v}$ are typical of variously populated groups with $M_{vir}/L >$ 40 $M_{\odot}/L_{\odot}$. The similarity of $M_{vir}/L$ and $S_{v}$ for these triplets to the corresponding parameters for more densely populated groups may stem from the fact that these triplets are located in clusters and/or have more distant companions, for example, as in Hickson groups where dwarf galaxies were found at distances of several group radii (Hunsberger et al. 1998). 

According to Coziol et al. (2004), the evolution level of a group increases with system mass and with galaxy velocity dispersion in the group and elliptical galaxies dominate in evolved groups. Tovmassian et al. (2004) also showed that the velocity dispersion in a group increases with number of $E/S0$ galaxies, with the frequency of occurrence of early-type galaxies being an increasing function of the group population and, accordingly, its mass. We found a similar dependence for the sample of interacting galaxies in groups, $\log S_{v} = 0.50 N(E/S0) + 1.96$, with the correlation coefficient R = 0.44 and the standard deviation SD = 0.36. However, the rms velocity of the group galaxies depends on the mean radial velocity, as can be seen from the linear regression equation: $\log S_{v} = 0.83 \log\langle V_{LG} \rangle - 1.03$, R = 0.43, and SD = 0.36. Note that these two dependences have equal correlation coefficients. Thus, the morphological classification of galaxies may be subject to distance selection, since the spiral arms in distant galaxies are more difficult to distinguish. 

Numerical simulations suggest (see, e.g., Zheng et al. (1993) and references therein) that the product of a galaxy merger is an elliptical galaxy. Observations of elliptical galaxies in compact Hickson groups showed that their properties differ from those of elliptical field galaxies (de Oliveira and Hickson 1994). The frequency of occurrence of $E/S0$ galaxies as a function of the population in interacting groups is 0.43, which is much higher than that in isolated CIG galaxies, 0.19. This result is acceptable for the merger hypothesis, especially if it is considered that these galaxies could have already undergone mergers in the past (Zheng et al. 1993; Hunsberger et al. 1998). This may be evidenced by the observational data on binuclear galaxies in compact groups, which could be the result of previous mergers (Amram et al. 2004; Bentoni and Buson 2000), and the presence of double nuclei in Markarian galaxies and their tendency to be found in pairs (Keel and van Soest 1992). \\

\section{Conclusion}
Based on the catalog by Vorontsov-Velyaminov (2001), we compiled a sample of 87 interacting galaxies with known radial velocities and with three to eight components. The median parameters for these galaxies are: the rms velocity $S_{v}$ = 166 km/s, the harmonic mean radius $R_{h}$ = 29 kpc, the system crossing time $\tau$ = 0.03 1/$H_{0}$, the virial mass $M_{vir}$ = 2.29 $\times  10^{12}$ $M_{\odot}$, the luminosity $L$ = 7.53 $\times 10^{10}$ $L_{\odot}$, and the mass-to-light ratio $M_{vir}/L$ = 33 $M_{\odot}/L_{\odot}$.

The interacting galaxies in groups and the single peculiar galaxies differ in morphological composition, as confirmed by their color indices: bluer galaxies are typical of the sample of single galaxies, while redder galaxies are typical of the galaxies in groups. The samples of interacting and peculiar galaxies contain about twice as many galaxies with UV than CIG galaxies.

The fraction of active galaxies in interacting groups correlates with the group population: the more components in a group, the fewer galaxies
with UV excess it contains (9\% in triplets, 7\% in quartets, and only 3\% active galaxies in groups of five to eight members). The frequency of occurrence
of $E/S0$ galaxies in interacting groups is 0.33, 0.50, and 0.67 for triplets, quartets, and groups with five to eight members, respectively. This is much higher than that in isolated CIG objects (0.19) and in peculiar galaxies (0.07), where the contribution of active galaxies is 4\% and 8\%, respectively. Thus, the contribution of early-type galaxies increases and the number of active galaxies decreases with increasing group population. The velocity dispersion of the group, its mass, luminosity, and virial mass-to-light ratio also increase with increasing group population. An excess of groups where all components are active galaxies with UV excess are observed in triplets. These results may suggest that interacting galaxies in differently populated groups are characterized by a certain morphological composition and are at different evolutionary stages.

\bigskip

{\bf ACKNOWLEDGMENTS.} 
I am grateful to V.E. Karachentseva and I.B.Vavilova for helpful discussions of the results and valuable
remarks that contributed to the improvement of this work. I used the LEDA (http://leda.univ-lyon1.fr), NED (http://nedwww.ipac.caltech.edu), and DSS images (http://archive.eso.org/dss/dss). 

{}
\end{document}